\documentclass[aps,prl,twocolumn,floatfix,showpacs]{revtex4}
\usepackage{amssymb}

\usepackage{epsfig}
\usepackage{amsmath}

\begin{document}

\title{A consistent interpretation of the low temperature magneto-transport in graphite\\using the  Slonczewski--Weiss--McClure 3D band structure calculations}
\author{J. M. \surname{Schneider}$^{}$}
\email{Johannes.Schneider@grenoble.cnrs.fr}
\author{M. \surname{Orlita}$^{}$}
\author{M. \surname{Potemski}$^{}$}
\author{D. K. \surname{Maude}$^{}$}
\affiliation{$^{}$Grenoble High Magnetic Field Laboratory, CNRS, 38042 Grenoble, France}

\date{\today}

\begin{abstract}
Magnetotransport of natural graphite and highly oriented pyrolytic graphite (HOPG) has been measured at mK
temperatures. Quantum oscillations for both electron and hole carriers are observed with orbital angular momentum
quantum number up to $N\approx90$. A remarkable agreement is obtained when comparing the data and the predictions of
the Slonczewski--Weiss--McClure tight binding model for massive fermions. No evidence for Dirac fermions is observed in
the transport data which is dominated by the crossing of the Landau bands at the Fermi level, corresponding to
$dE/dk_z=0$, which occurs away from the $H$ point where Dirac fermions are expected.
\end{abstract}


\maketitle

Recently, massless Dirac fermions have been observed at the $K$ point of the Brillouin zone in graphene, a hexagonally
arranged carbon monolayer with quite extraordinary properties~\cite{Novoselov05}. Historically, graphene forms the
starting point for the Slonczewski, Weiss and McClure (SWM) band structure calculations of
graphite~\cite{Slonczewski58,McClure60}. In graphite, the Bernal stacked graphene layers are weakly coupled with the
form of the in-plane dispersion depending upon the momentum $k_z$ in the direction perpendicular to the layers. The
carriers occupy a region along the $H-K-H$ edge of the hexagonal Brillouin zone. At the $K$ point ($k_z=0$), the
dispersion of the electron pocket is parabolic (massive fermions), while at the $H$ point ($k_z=0.5$) the dispersion of
the hole pocket is linear (massless Dirac fermions). A clear signature of Dirac fermions at the $H$ point of graphite
has recently been reported using far-infrared magneto-absorption measurements~\cite{Orlita08}. Such measurements probe
the very close vicinity of the $H$ and $K$ points where there is a maximum in the joint density of states.

The SWM model, which provides a remarkably accurate description of the band structure, has been extensively tested
using Shubnikov de Haas, de Haas van Alphen, thermopower and magneto-reflectance measurements to caliper the Fermi
surface of graphite~\cite{Soule58,Soule64,Woollam70,Woollam71a,Williamson65,Schroeder68}. There are even reports of a
charge density wave state above $B=22$~T~\cite{Iye82,Timp83,Iye85}. However, the observation of massless carriers with
a Dirac--like energy spectrum, using magneto-transport measurements~\cite{Luk04,Luk06} remains controversial, since in
the SWM model, the electrons and hole carriers at the Fermi level are both massive quasi--particles.

In this Letter, we report magneto-transport measurements of natural graphite at very low temperature ($T\approx
10$~mK). Due to the low temperatures used, the magneto-transport is much richer than previously published
data~\cite{Soule58,Soule64,Woollam70,Woollam71a,Williamson65,Iye82,Timp83,Iye85,Luk04,Luk06}. Quantum oscillations are
observed for both majority electrons and holes with orbital quantum number up to almost N=100. We show that these
oscillations are fully consistent with the presence of majority electron and hole pockets within the three dimensional
SWM band structure calculations for graphite. At high magnetic fields ($B>2$~T), a significant deviation from 1/B
periodicity occurs due to the well documented movement of the Fermi energy as the quantum limit is
approached~\cite{Sug66,Woollam71a}. This seriously questions the validity of using the high field data to extract the
phase of the Shubnikov de Haas oscillations, and hence the nature of the charge carriers~\cite{Luk06}.

For the measurements mm-size pieces of natural graphite and highly oriented pyrolytic graphite (HOPG), a few hundred
microns thick, where contacted in an approximate Hall-bar configuration using silver paint. The measurements were
performed with the sample placed directly in the mixture of a He$^3$/He$^4$ dilution fridge, using an \textit{ac}
current of $\sim10$~$\mu$A at $10.7$~Hz and conventional phase sensitive detection. The magnetic field was applied
along the c--axis of the sample.

\begin{figure}[]
\begin{center}
\includegraphics[width= 8.5cm]{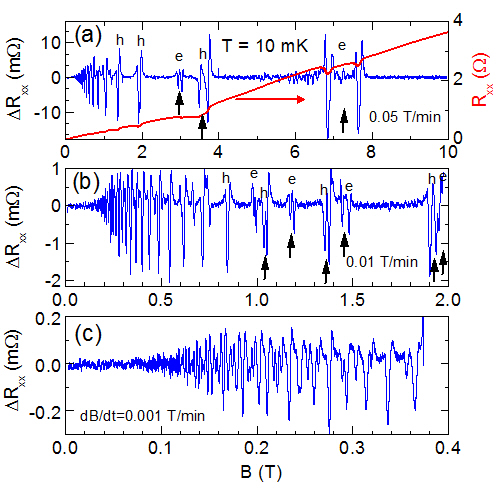}
\end{center}
\caption{(color online) (a) Right axis: Resistance $R_{xx}$ versus B measured at $T=10$~mK for natural graphite. (a-c)
Left axis: Background removed data $\Delta R_{xx}$ showing quantum oscillations measured over different magnetic field
regions. The high field electron (e) and hole (h) features are indicated. The vertical arrows indicate spin split
electron and hole features.} \label{Fig1}
\end{figure}

Typical low temperature data for $R_{xx}$~$\mathrm{(\Omega)}$ as a function of the magnetic field from $B = 0-10$~T for
natural graphite, is shown in Fig.~\ref{Fig1}(a). $R_{xx}(B)$ increases roughly linearly with the magnetic field and at
$B = 10$~T, it is about three orders of magnitude larger than the zero-field
value~\cite{McClure68,Abrikosov99,Du05,Gonz07}. In addition, quantum oscillations are superimposed on the large
magneto-resistance background. These oscillations, can be better seen in the background removed data $\Delta R_{xx}$
plotted in Fig.~\ref{Fig1}(a-c) for successively slower sweeps in order to reveal the quantum oscillations in the
different magnetic field regions. The background can either be removed by subtracting a smoothed (moving window
average) data curve or by numerically calculating the second derivative $d^2R/dB^2$. Both techniques give similar
results and here we use averaging to remove the background. As the oscillations are periodic in $1/B$, the optimal
number of points used in the averaging depends upon the magnetic field region. For this reason, the amplitudes of the
oscillations in Fig.~\ref{Fig1}(a-c) should not be compared, as different averaging was used to remove the background.
HOPG (not shown) presents almost identical oscillations, with very slightly different frequencies, and a significantly
reduced amplitude~\cite{OrlitaJP08}. For this reason we concentrate here on the data for natural graphite.

\begin{figure}
\begin{center}
\includegraphics[width=8.5cm]{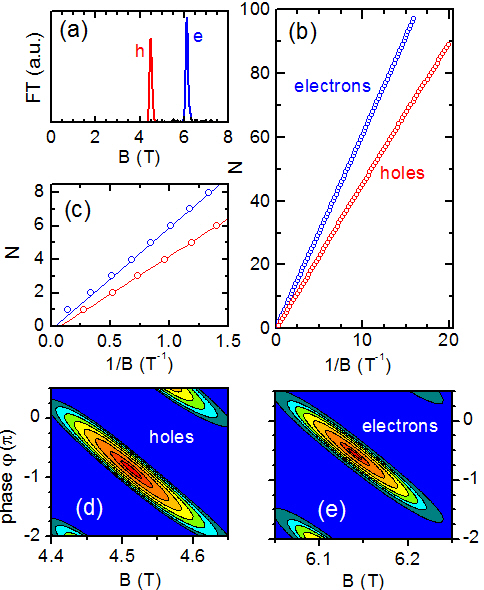}
\end{center}
\caption{(color online) (a) Fourier transform of the low magnetic field $\Delta R_{xx}(1/B)$. (b-c) Orbital angular
momentum quantum number N, as a function of the reciprocal magnetic field positions of the electron and hole features.
(d) and (e) Contour plot of the phase shift function $K(\varphi,B)$ in the vicinity of the hole and electron features.
Maxima in $K(\varphi,B)$ determines the frequency and phase of the oscillations.} \label{Fig2}
\end{figure}

In the $\Delta R_{xx}(B)$ data shown in Fig.~\ref{Fig1} two series of oscillations can be distinguished. The
oscillations start at a magnetic field $B\approx0.1$~T, and spin splitting of the features (indicated by arrows) is
observed for magnetic fields $B>1$~T, compared to previous work~\cite{Woollam70} in which spin-splitting was only
observed for the three last features at magnetic fields $B>2$~T. The electronic g-factor $g_s$ can be estimated from
the magnetic fields at which spin splitting occurs, ($B_{z}\sim 1$~T), and at which the Shubnikov de Haas oscillations
start ($B_{c}\sim 0.07$~T). For a Landau level broadening $\Gamma$, spin splitting occurs when
$g_s\mu_{B}B_z\approx\Gamma$, and Shubnikov de Haas oscillations occur when $\hbar e B_c/m^*\approx\Gamma$ where
$m^*=0.056 m_e$ is the electron effective mass~\cite{Nozieres58}. Assuming $\Gamma$ to be field independent we can
write $g_s\approx\hbar e B_c/m^* \mu_B B_z\approx2.5$. The mobility, estimated from the condition $\mu B_c=1$, is $\sim
140,000$~cm$^2$/Vs. In the Fourier transformation of the $0-0.4$~T $\Delta R_{xx}$ versus $(1/B)$ data, shown in
Fig.~\ref{Fig2}(a), two frequencies are found and assigned to the electron pocket at the $K$ point ($B_{F_e}=6.14$~T)
and hole pocket at the $H$ point ($B_{F_h}=4.51$~T). This assignment is the well established in the
literature~\cite{Woollam70,Woollam71a} and it is the only assignment which is consistent with the magnetoreflectance
measurements~\cite{Schroeder68}. For HOPG (data not shown) we find slightly higher frequencies, $B_{F_e}=6.49$~T and
$B_{F_h}=4.73$~T.

In graphite, we have experimentally $\rho_{xx}\gg\rho_{xy}$ so that the tensor relation for conductivity simplifies to
give $\sigma_{xx}\propto \rho_{xx}^{-1}$. Therefore, conductivity maxima which occur at coincidence of Landau bands and
the Fermi energy $E_F$~\cite{Adams58}, correspond to minima in $\Delta R_{xx}(B)$. We perform a classical 1/B analysis
of our data assigning an orbital quantum number N to the electron and hole minima of $\Delta R_{xx}(B)$. For $N<25$ the
magnetic field positions of each series of oscillations can be determined directly from $\Delta R_{xx}(B)$. For $N>25$,
pass band frequency domain filtering was used to separate the superimposed electron and hole features. The position of
the features in inverse magnetic field versus N is shown in Fig.~\ref{Fig2}(b). For both electrons and holes we can see
features with angular quantum number $1<N<90$ (to almost $N=100$ for electrons). $N$ versus $1/B$ has a linear
dependence and the slope gives the fundamental fields $B_{F_e}=6.10\pm0.05$~T and $B_{F_h}=4.50\pm0.05$~T, in good
agreement with the values obtained from the Fourier transform.

At low magnetic fields, i.e. at high quantum number N, a perfect linear behavior in $N(1/B)$ is observed for both
electrons and holes. For high magnetic fields, i.e. for low $N$, clear deviations from the linear behavior are observed
for the electron features (see Fig.~\ref{Fig2}(c)). This deviation from a periodic in 1/B behavior at high magnetic
fields is due to the Fermi level moving as the quantum limit is approached in graphite~\cite{Sug66,Woollam71a}.
Clearly, the high field data should not be used to extract the phase of the oscillations~\cite{Luk06}. Equally,
extrapolating the low field data to find the intercept does not give a reliable estimate of the phase. Instead, we
prefer to use the phase shift analysis method developed by Luk'yanchuk and Kopelevich~\cite{Luk04}, to extract the
phase from the complex Fourier transform $\hat{f}(B)$ of the low magnetic field $\Delta R_{xx}(1/B)$. The phase shift
function $K(\varphi,B)=Re[e^{-i\varphi}\hat{f}(B)]$ has maximum in the $\varphi-B$ plane which can be used to extract
both the frequency ($B$) and phase ($\varphi$) of the oscillations. $K(\varphi,B)$ is plotted in Fig.~\ref{Fig2}(d-e)
in the regions of the hole and electron features. From the maxima, the determined frequency and phase are
$B_{f_h}=4.51$~T, $\varphi_h=-(0.56\pm0.1)\pi$ and $B_{f_e}=6.14$~T, $\varphi_e=-(0.86\pm0.1)\pi$ for the hole and
electron features respectively. For HOPG a similar analysis gives $\varphi_h=-(1.04\pm0.1)\pi$ and
$\varphi_e=-(0.92\pm0.1)\pi$.

The oscillatory conductivity can be written $\Delta \sigma \propto \cos(2\pi B_{f}/B-2\pi\gamma+\delta)$ with
$\gamma=1/2$ for massive fermions and $\gamma=0$ for massless Dirac fermions~\cite{Adams58,Mikitik06}. At low magnetic
fields inter Landau level scattering is expected to dominate so that $\delta=\pi/4$ for a 3D corrugated Fermi surface
($\delta=0$ for a 2D cylindrical Fermi surface). The expected value of the phase $\varphi=-2\pi\gamma+\delta$ for
massive 3D fermions ($\gamma=1/2$) is therefore $\varphi=-0.75\pi$ in reasonable agreement with the experimental phase
for both electrons and holes. For HOPG the phase is also consistent with $\gamma=1/2$ but with $\delta\approx0$. The
value of $\gamma=1/2$ is in agreement with published results~\cite{Soule64,Williamson65} and theoretical
considerations~\cite{Mikitik06}. In contrast, the prediction for 2D massless Dirac fermions ($\gamma=0$) with
$\varphi=0$ is completely inconsistent with the determined phase for both electrons and holes. We therefore conclude
that there is no evidence from transport measurements for the existence of masseless Dirac fermions with a Berry phase
$\gamma=0$. Nevertheless, there is compelling evidence from far-infrared absorption, for the existence of Dirac
fermions at the $H$ point in graphite~\cite{Orlita08}. Far infrared measurements probe carriers in the very close
vicinity of the $H$ point where there is a maximum in the joint (initial and final) density of states. Transport
measurements however, are sensitive to the density of states at $E_F$, which is modulated with increasing magnetic
field, as the Landau bands cross the Fermi energy. For holes, maxima in the density of states correspond to Landau
bands crossing $E_F$ for $k_z<0.5$, away from the $H$ point, where the dispersion is no longer linear and \emph{a
priori} there is no reason to expect the carriers to behave as Dirac fermions.

\begin{figure}
\begin{center}
\includegraphics[width= 8.5cm]{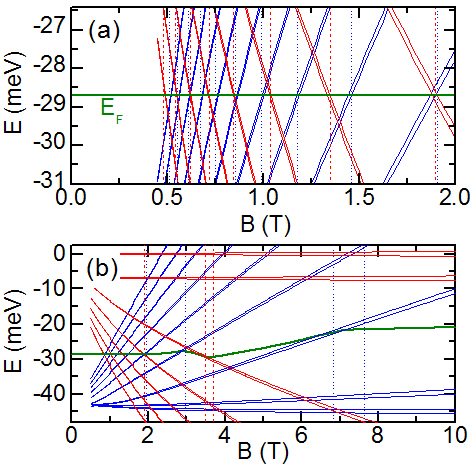}
\end{center}
\caption{(color online) (a) Electron (increasing with B) and hole (decreasing with B) Landau bands ($dE/dk_z=0$) as a
function of the magnetic field (calculated from SWM model for $B\geq0.4$~T). The crossing of the electron and hole
bands with the Fermi energy is in nearly perfect agreement with the measured magnetic field position of the electron
(dotted lines) and hole (dashed lines) features. b) For $B\geq2$~T, the Fermi energy is calculated self consistently,
keeping the sum of the electron and hole concentrations constant.} \label{Fig3}
\end{figure}

It is interesting to compare the data with the predictions of the SWM 3D band structure model with its seven tight
binding parameters $\gamma_0,...,\gamma_6$. When the parameter $\gamma_3$ is taken into account the magnetic field
Hamiltonian has infinite order. This was numerically reduced to a $600\times600$ matrix for the exact diagonalization
procedure. A maximum in the conductivity is expected when there is a maximum in the density of states at the Fermi
level. This can be found, for a given magnetic field, by looking for a Landau band with a slope $dE/dk_z=0$ at $E_F$,
which we refer to as a Landau band `crossing' the Fermi level. Using the tight binding parameters of
Ref.~\cite{Brandt88} as a starting point, this procedure was repeated until a satisfactory agreement with the data was
obtained. The tight binding parameters found are given in Table~\ref{tab1}. While we are unable to fit our data with
exactly the same tight binding parameters as in Ref.~\cite{Brandt88}, the values we find are nevertheless not
significantly different. Moreover, the predicted~\cite{Koshino07} effective mass, $m^*=4\hbar^2\gamma_1/3
a_0^2\gamma_0^2=0.054m_e$ where $a_0=0.246$~nm is the in-plane lattice constant, calculated using our values for
$\gamma_0$ and $\gamma_1$, is in good agreement with the accepted value~\cite{Nozieres58}.

\begin{table}
\begin{center}
\begin{tabular}{c|c|c}
& This work & Ref.~\cite{Brandt88}\\ 
\hline
$\gamma_0$~(eV) & $3.37 \pm 0.02$ & $3.16 \pm 0.05$\\
$\gamma_1$~(eV) & $0.363 \pm 0.05$ & $0.39 \pm 0.01$\\
$\gamma_2$~(eV) & $-0.0243 \pm 0.001$ & $-0.02 \pm 0.002$\\
$\gamma_3$~(eV) & $0.31 \pm 0.05$ & $0.315 \pm 0.015$\\
$\gamma_4$~(eV) & $0.07 \pm 0.01$ & $0.044 \pm 0.024$\\
$\gamma_5$~(eV) & $0.05 \pm 0.01$ & $0.038 \pm 0.005$\\
$\gamma_6=\Delta$~(eV)   & $-0.007$ & $-0.008 \pm 0.002$\\
$E_F$~(eV)    & $-0.0287$ & $-0.024 \pm 0.002$\\
$g_s$      & $2.4\pm 0.1$ & -\\
$n_0(\mathrm{cm}^{-3})$      &  $-(2.4 \pm 0.4)\times 10^{17}$ &  -\\
\end{tabular}
\end{center}
\caption{Summary of the SWM tight binding parameters found in this work and compared to the values given in
Ref.~\cite{Brandt88}.}\label{tab1}
\end{table}

For magnetic fields below $2$~T it is a good approximation to assume that the Fermi level is constant. The electron and
hole Landau bands (solid lines), calculated using the parameters in Table~\ref{tab1}, are plotted in Fig.~\ref{Fig3}(a)
for magnetic fields below $2$~T. The vertical broken lines indicate the observed electron (dotted) and hole (dashed)
minima in $\Delta R_{xx}$ (maxima in $\sigma_{xx}$). The agreement between the magnetic field position of the Landau
bands crossing the Fermi level and the features in the transport data is remarkable.

At higher magnetic fields, as graphite approaches the quantum limit, the Fermi energy is no longer constant as carriers
are transferred between the electron and hole pockets. This is the reason for the considerable deviation from 1/B
periodicity observed at high magnetic fields in Fig.~\ref{Fig2}(c). Nevertheless, as can be seen in Fig.~\ref{Fig3}(b),
the SWM model can correctly predict the magnetic field position of the features provided the movement of $E_F$ is taken
into account. Here the Fermi level has been calculated self-consistently assuming the sum of the electron and hole
concentrations is constant, $n-p=n_0$. The electron concentration corresponds to the number of states in partially
filled bands below the Fermi energy, and the hole concentration to those above the Fermi energy. In order to fit the
low--field data, we have used $n_0=-2.4\times 10^{17}$~$\mathrm{cm}^{-3}$, assuming that under neutrality conditions,
$n=p=8\times 10^{18}$~$\mathrm{cm}^{-3}$, with $n=64\gamma_1|\gamma_2|/(9\sqrt{3}\pi^2\gamma_0^2a_0^2c_0)$
\cite{Brandt88}.

To reproduce the spin splitting in the high magnetic field data, a g-factor $g_s=2.4$ is required. In graphite, the
g-factor cannot be reliably estimated from the separation $\Delta B$ of the spin split features since the Fermi energy
is moving with field~\cite{Woollam70}. While this does not noticeably shift the orbital features below $B=2$~T, the
shift is significant compared to the spin gap. The value $g_s=2.4$ should be considered as a lower limit. Any
significant Landau level broadening, neglected in our model, would reduce the movement of the Fermi energy, and
therefore increase the value of $g_s$ required to fit the data.

To conclude, low temperature magnetotransport data of natural graphite and HOPG can be fully explained using the
Slonczewski--Weiss--McClure tight binding model for massive fermions. No evidence for Dirac fermions at the $H$ point
is observed in the transport data. This can be understood, since transport is dominated by the crossing of the Landau
bands at the Fermi level, corresponding to $dE/dk_z=0$, which occurs away from the $H$ point ($k_z=0.5$), where the
carriers are indeed Dirac fermions~\cite{Gruneis08,Orlita08}.

\acknowledgements{This work has been partially supported by ANR contract PNANO-019-06. We acknowledge useful
discussions with I.A. Luk'yanchuk and Y. Kopelevich.}


\end{document}